\documentstyle[aps]{revtex}
\begin{document}
\draft
\input{epsf}

\title{ Production of $^{3}_{\Lambda}H$ and $^{4}_{\Lambda}H$ in Central 11.5
 GeV/c Au + Pt Heavy Ion Collisions}

\author{
T.A. Armstrong                \unskip,$^{(8,a)}$
K.N. Barish                   \unskip,$^{(3)}$
S. Batsouli                   \unskip,$^{(13)}$
S.J. Bennett                  \unskip,$^{(12)}$
M. Bertaina                   \unskip,$^{(7,b)}$
A. Chikanian                  \unskip,$^{(13)}$
S.D. Coe                      \unskip,$^{(13,c)}$
T.M. Cormier                  \unskip,$^{(12)}$
R. Davies                     \unskip,$^{(9,d)}$
C.B. Dover                    \unskip,$^{(1,e)}$
P. Fachini                    \unskip,$^{(12,o)}$
B. Fadem                      \unskip,$^{(5)}$
L.E. Finch                    \unskip,$^{(13)}$
N.K. George                   \unskip,$^{(13,f)}$
S.V. Greene                   \unskip,$^{(11)}$
P. Haridas                    \unskip,$^{(7,g)}$
J.C. Hill                     \unskip,$^{(5)}$
A.S. Hirsch                   \unskip,$^{(9)}$
R. Hoversten                  \unskip,$^{(5)}$
H.Z. Huang                    \unskip,$^{(2)}$
H. Jaradat                    \unskip,$^{(12)}$
B.S. Kumar                    \unskip,$^{(13,h)}$
T. Lainis	              \unskip,$^{(10)}$
J.G. Lajoie                   \unskip,$^{(5)}$
R.A. Lewis                    \unskip,$^{(8)}$
Q. Li                         \unskip,$^{(12)}$
B. Libby                      \unskip,$^{(5,i)}$
R.D. Majka                    \unskip,$^{(13)}$
T.E. Miller                   \unskip,$^{(11)}$
M.G. Munhoz                   \unskip,$^{(12)}$
J.L. Nagle                    \unskip,$^{(4)}$
I.A. Pless                    \unskip,$^{(7)}$
J.K. Pope                     \unskip,$^{(13,j)}$
N.T. Porile                   \unskip,$^{(9)}$
C.A. Pruneau                  \unskip,$^{(12)}$
M.S.Z. Rabin                  \unskip,$^{(6)}$
J.D. Reid                     \unskip,$^{(11,k)}$
A. Rimai                      \unskip,$^{(9,l)}$
A. Rose                       \unskip,$^{(11)}$
F.S. Rotondo                  \unskip,$^{(13,m)}$
J. Sandweiss                  \unskip,$^{(13)}$
R.P. Scharenberg              \unskip,$^{(9)}$
A.J. Slaughter                \unskip,$^{(13)}$
G.A. Smith                    \unskip,$^{(8)}$                                                 X
M.L. Tincknell                \unskip,$^{(9,n)}$
W.S. Toothacker               \unskip,$^{(8,e)}$
G. Van Buren                  \unskip,$^{(7,2,o)}$
F.K. Wohn                     \unskip,$^{(5)}$
Z. Xu                         \unskip$^{(13)}$}
\address{\centerline{(The E864 Collaboration)}}
\address{  $^{(1)}$ Brookhaven National Laboratory, Upton, 
New York 11973 \break
  $^{(2)}$ University of California at Los Angeles, Los Angeles, 
California 90095 \break  
  $^{(3)}$ University of California at Riverside, Riverside, 
California 92521 \break
  $^{(4)}$ Columbia University, New York 10027 \break
  $^{(5)}$ Iowa State University, Ames, Iowa 50011 \break 
  $^{(6)}$ University of Massachusetts, Amherst, Massachusetts 01003 \break 
  $^{(7)}$ Massachusetts Institute of Technology, Cambridge, 
Massachusetts 02139 \break 
  $^{(8)}$ Pennsylvania State University, University Park, 
Pennsylvania 16802 \break 
  $^{(9)}$ Purdue University, West Lafayette, Indiana 47907 \break 
  $^{(10)}$ United States Military Academy, West Point, New York 10996 \break
  $^{(11)}$ Vanderbilt University, Nashville, Tennessee 37235 \break 
  $^{(12)}$ Wayne State University, Detroit, Michigan 48201 \break 
  $^{(13)}$ Yale University, New Haven, Connecticut 06520 \break
}

\date{\today}
\maketitle
\begin{abstract}
We present measurements from BNL AGS Experiment E864 of the $^{3}_{\Lambda}H$ 
invariant multiplicity and of the 90\% Confidence Level upper limit on the 
$^{4}_{\Lambda}H$ yield in central 11.5 A GeV/c Au~+~Pt collisions.  
The measurements span a rapidity range from center
of mass, $y_{cm}$, to $y_{cm}$+1 and a transverse momentum range of
 $0.< p_{T}\le 1.5$ GeV/c. We compare these results with
E864 measurements of stable light nuclei 
and particle unstable nuclei yields of the same baryon number.  The
implications of these results for the coalescence of strange clusters are discussed.
\end{abstract}

\section{Introduction}
Relativistic heavy ion collisions are the main experimental tool for studying 
the
behaviour of nuclear matter under extreme energy and baryon density conditions.
In addition, these collisions offer the only method to produce
multistrange bound systems in a controlled manner, since there is copious
 strangeness production in these collisions.

Hypernuclei, which are nuclei where a nucleon is replaced by a $\Lambda$
hyperon, exist and have been studied for many years. More exotic forms of multi-strange
nuclear systems have been hypothesized to exist. These include both hypernuclei which
contain more than one hyperon, so called MEMOS (Metastable Exotic Hypernuclear Objects)
\cite{sch} and strangelets \cite{wit,jaf1,jaf2} which are single ``bags'' of
approximately equal numbers of strange, up and down quarks. In many cases the quantum
numbers of the proposed MEMOS and of strangelets are the same. If so, and if
strangelets exist, the MEMOS would decay into the more deeply bound strangelets . The
production of these exotic hypernuclei could then be a doorway to the production of
strangelets.
 In Experiment 864, 30 billion $10\%$ most central,
 11.5 GeV/c per nucleon Au on Pt or Pb collisions
were sampled for the search for strangelets with $A<100$ and lifetimes greater
 than 50 ns. No strangelets were observed at a level of $\approx 10^{-8}$ per
central collision \cite{arm1,arm2,arm3}. 

The study of the production of the light hypernuclei $^{3}_{\Lambda}H$ and 
$^{4}_{\Lambda}H$ is essential 
in understanding the production mechanism of exotic objects such 
as multi-hypernuclei or MEMOS and the strangelets they
might decay into if the latter are more stable. There are various possible 
production mechanisms for multihypernuclei and strangelets such as the QGP
distillation scenario \cite{liu,cra,grei}, coalescence mechanisms 
\cite{sch,bal} and thermal models \cite{brau,brau2,raf,raf2}. When a number of
nucleons coalesce to form a nucleus there is a 
``penalty factor'' involved in the yields of the various nuclei
 for each nucleon that is added to a cluster. In the
case of hypernuclei there is also an additional suppression factor due to the
different yields of strange baryons as compared with protons and neutrons and
there might also be an extra ``strangeness penalty factor'' for coalescing
 strange particles. 
The study of light nuclei in E864 \cite{arm4}
is informative about the coalescence process of nucleons at freeze-out
 and the
``penalty factor'' involved when adding a nucleon to a cluster.
When the invariant yields of light nuclei with
A=1 to A=7 are examined in a small kinematic region near the center of mass rapidity and at low $p_{T}$ 
($p_{T}/A \leq 300 MeV$), they show an
exponential dependence with baryon number, suggesting a penalty factor of
 approximately 48 for each nucleon
added \cite{arm4}. However in order to determine whether there is some
extra ``strangeness penalty factor'' when hyperons are coalesced, the study of 
$^{3}_{\Lambda}H$ and $^{4}_{\Lambda}H$ is important.
Finally the production of $^{3}_{\Lambda}H$ and $^{4}_{\Lambda}H$ in
 relativistic heavy
ion collisions is a novel measurement and interesting in itself for
 understanding the
strangeness degree of freedom in hadronic systems. In this paper we  
present measurements from BNL AGS Experiment E864 of the $^{3}_{\Lambda}H$ 
invariant multiplicity and of the 90\% Confidence Level upper limit on the 
$^{4}_{\Lambda}H$ yield.

\section{The E864 Spectrometer}

\begin{figure}
\centering\leavevmode\epsfbox{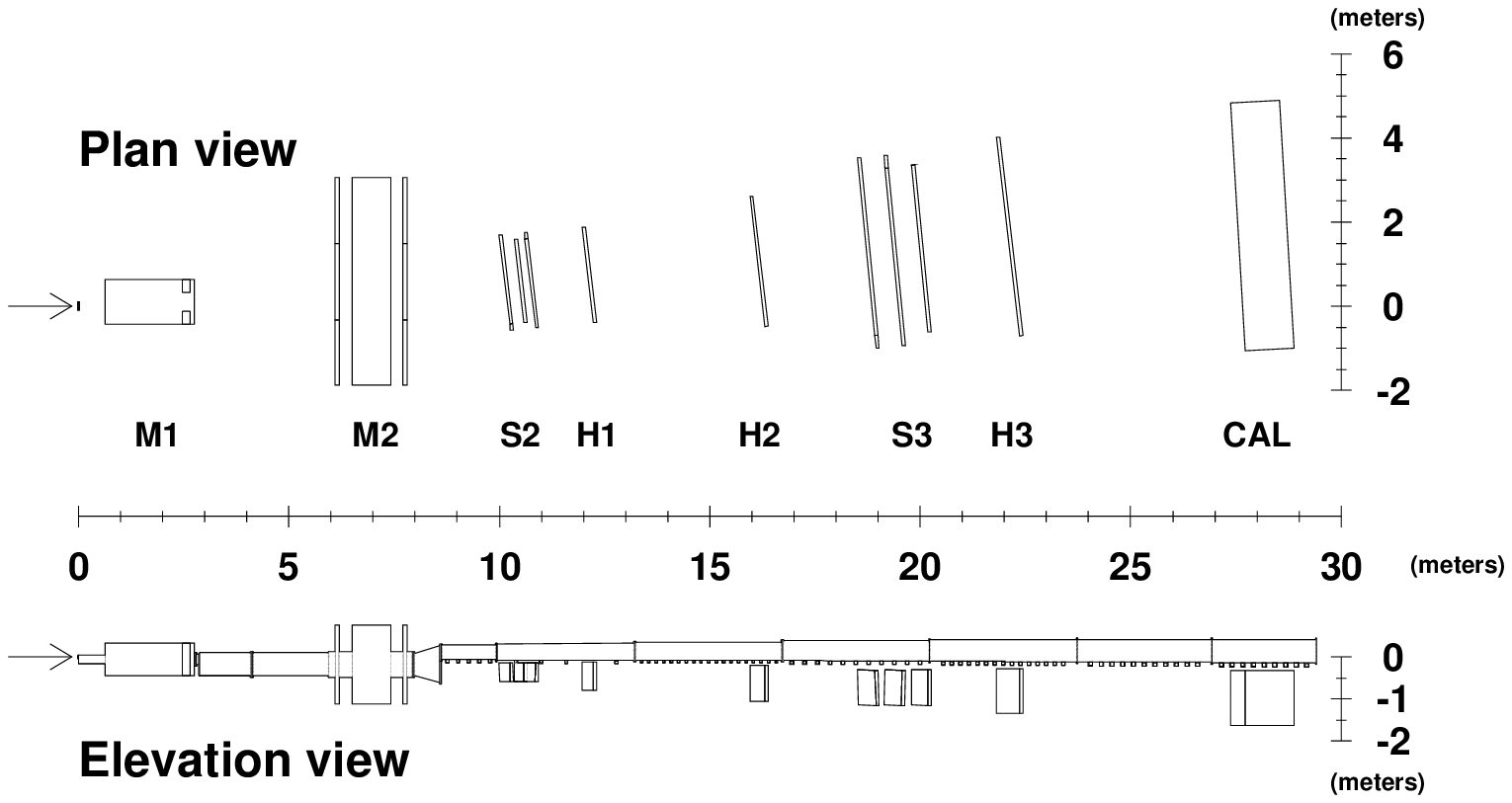}
\caption{The E864 spectrometer in plan and elevation views, showing the
dipole magnets (M1 and M2), hodoscopes (H1, H2, and H3), straw tube arrays 
(S2 and S3) and hadronic calorimeter (CAL).  The vacuum chamber is not 
shown in the plan view. }
\label{fig:john_ap}
\end{figure}

Experiment 864 is an open geometry, high data rate spectrometer 
designed primarily for the search of strange quark matter produced 
in relativistic Au + Pt collisions.  The open geometry allows for
 a large
region of the produced phase space for heavy clusters to be sampled. A beam of Au ions
 with momentum 
11.5 GeV/c per nucleon is incident on a fixed Pt target. 
 The interaction products
can be identified by their charge and mass in the tracking system and by their
energy and time of flight
in the calorimeter. A detailed description of the E864 apparatus is given in
 \cite{bignim}. Diagrams of the plan and elevation views of the apparatus are
 given in Figure~\ref{fig:john_ap}.

The tracking system consists of two dipole analyzing magnets (M1 and M2) with vertical fields, three
scintillator time of flight hodoscope planes (H1, H2 and H3) and two straw-tube stations
 (S2 and S3).
The dipole magnets M1 and M2 
can be set to different field strengths to optimize the acceptance for
various particles of interest. The three hodoscope planes measure the time,
 charge and spatial position for
each charged particle that passes through them.  
The straw tube planes provide improved spatial resolution for the charged
 tracks. 
The magnetic rigidity, momentum, and mass of the tracked particles 
can be determined by the position, time, and charge information 
from these detectors together with the knowledge of the magnetic field and the
assumption that they come from the target.
 The hadronic calorimeter measures
 the energy 
and time of flight for all particles. The calorimeter is the primary
detector for identifying neutral particles and can act as a powerful tool for
background rejection for charged particles. It has excellent resolution for 
hadronic showers in energy ($\sigma_{E}/E = .34/(\sqrt{E}) + .035$ for $E$ in
units of GeV) and time ($\sigma_{t} \approx$ 400 ps) and is described 
in detail in \cite{claudecal}. There is a vacuum tank along the beam line
to reduce the background from beam particles interacting downstream.
Just upstream of the target there are beam counters and a multiplicity
 counter
which are used to set a first level trigger that selects interactions
 according to
their centrality. The calorimeter energy and time of flight measurements
 are also used
to make a level-2 trigger (LET) that rejects interactions with no high
 mass particle in the spectrometer \cite{hil}.
                    
For this study the magnetic field of the
spectrometer was -0.2 Tesla, which is the optimum magnetic field for the
simultaneous acceptance of
both the decay products of the hypernuclei ($\pi^{-}$ and $^{3}He$ or 
$^{4}He$ ). We triggered on the $10\%$ most
central events
as defined by our multiplicity counters and an additional high mass LET
trigger which was set for the enhancement of 
$^{3}He$ and $^{4}He$ ions. The LET trigger rejected interactions that did not
result in any
high mass objects in the calorimeter by a factor of approximately 60. In this
 way $13.5\times 10^{9}$ 10\% most central collisions were sampled.

\section{Data analysis}
The light hypernuclei $^{3}_{\Lambda}H$ and $^{4}_{\Lambda}H$ 
decay weakly into mesonic and non-mesonic
channels. Their lifetimes of $\approx 2 \times10^{-10}sec$
  (c$\tau \approx $ 6 cm) 
 imply that they decay far outside the collision fireball, so we can observe them through
the detection of their decay products. The combination of our knowledge of 
the magnetic field and information obtained by the tracking detectors of the
 charge,
 time of flight and rigidity of the decay products can provide us with 
accurate mass
reconstruction of these particles \cite{arm4,bignim}.
Since E864 has good particle
identification for charged particles we have concentrated on the following
 mesonic channels
for the $^{3}_{\Lambda}H$ and $^{4}_{\Lambda}H$ :
\begin{description} 
\item [] $^{3}_{\Lambda}H \rightarrow  \pi^{-} + ^{3}He$,  branching ratio: 25\% \cite{kam,glo}
\end{description} 
\begin{description}
\item [] $^{4}_{\Lambda}H \rightarrow  \pi^{-} + ^{4}He$,  branching ratio: 50\% \cite{kum,out}.
\end{description} 

\subsection{$^{3}_{\Lambda}H$ analysis}
In order to be able to reconstruct the $^{3}_{\Lambda}H$ signal from its decay products
the ``mixed
event method'' is used, as explained below. First, events that contain at least one pair of the
decay products of $^{3}_{\Lambda}H$ are selected. The invariant                 
mass spectrum of the $^{3}He$ nuclei and the negative pions of each such event
is constructed. This ``same event'' invariant mass spectrum (SE) contains a small $^{3}_{\Lambda}H$ signal and a
large background that is due to particles that are not the decay products 
of a hypernucleus. The background shape (Bg) is obtained
by constructing the invariant mass spectrum of uncorrelated $^{3}He$, $\pi^{-}$ that come from
different events that contain at least one pair of the particles of interest. Specifically we combine the daughter
particle of one type from one event with the daughter particles of the other type from
a number of subsequent events. We have to make sure however that the mixed event spectrum
 does not contain pairs of overlapping $^{3}He$
and $\pi^{-}$, which could not be found if both tracks were in the same event. 
This is achieved by requiring that the $^{3}He$ and $\pi^{-}$ are in different
 sides of the detector horizontally (magnetic bend direction), with the sides assigned to give optimum 
efficiency for simulated decays. 
We then simulate
the shape of the hypernucleus mass peak by using a GEANT simulation of the decay products of 
$^{3}_{\Lambda}H$ passing through the apparatus and reconstructing 
their invariant mass spectrum (MC).
Finally we fit a linear combination of the Monte Carlo shape of the signal and 
the mixed event spectrum to the same event spectrum,   
\begin{equation}
\alpha\times (Bg)  +  \beta\times (MC)  =  SE  .
\label{eq:let_cor}
\end{equation}
The determination of the parameters $\alpha,\beta$ allow us to measure the signal
either by subtracting $\alpha\times (Bg)$ from the (SE) spectrum and then integrating the
subtracted spectrum over the region of the expected signal or as $\beta\times (MC)$.  

\begin{figure}
\centering\leavevmode\epsfbox{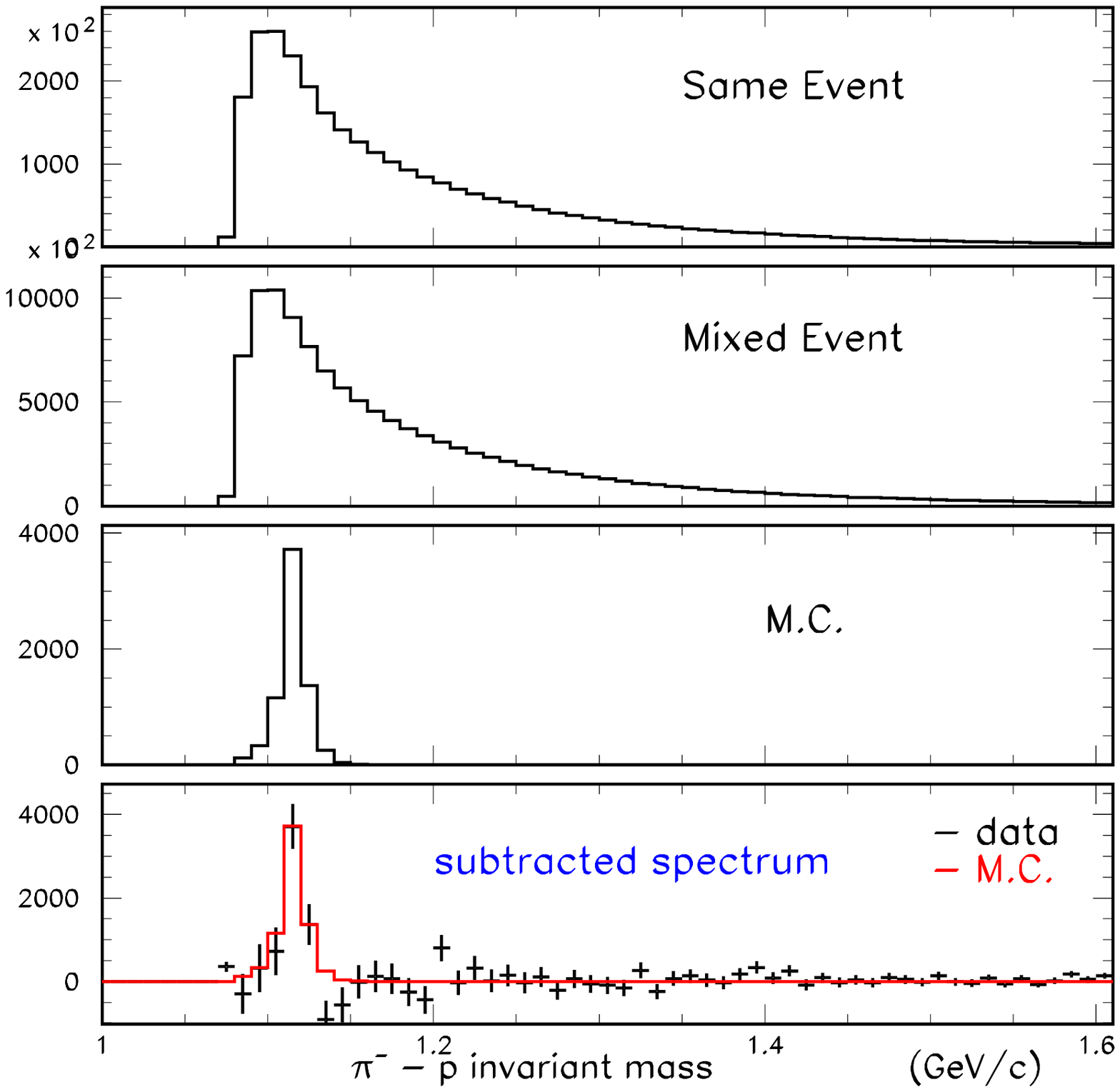}
\caption{The top panel shows the invariant mass spectrum for p - $\pi^{-}$ coming from
the same event. The second panel shows the invariant mass spectrum for p - $\pi^{-}$ 
coming from different events. The following panel shows the M.C. $\Lambda$ signal. The
bottom panel has the
subtracted invariant mass spectrum for p - $\pi^{-}$. The solid histogram overlaid on
the data is
the M.C. $\Lambda$ signal }
\label{fig:lam}
\end{figure}

As a check of this technique we looked at the p - $\pi^{-}$ invariant mass
spectrum from 6.5 million events
to observe the signal from lambda decays.
In Figure~\ref{fig:lam}  we show the same event and mixed event spectra, 
the M.C. signal and
the subtracted spectrum on which we overlay the M.C. signal. The agreement of the
M.C. and the data is good and the fit suggests a signal of 7.25 $\sigma$ above background. 
We have calculated invariant multiplicities of the $\Lambda$ in several rapidity and
transverse momentum bins and they are found to be in good agreement with measurements from
 E891 and E877 \cite{ahm,barr}.

In order to reconstruct the $^{3}_{\Lambda}H$ signal, we have to identify its
decay products. A $^{3}He$ track is defined as a charge two track
with rapidity less than 2.7 and reconstructed mass in the range
 $1. ~-~ 3.4$ $GeV/c^{2}$. The
efficiency of the rapidity cut is incorporated in the geometric acceptance for the 
$^{3}_{\Lambda}H$ and the efficiency of the mass cut is calculated to be 97\% 
by using real data. A systematic error of 1\% is associated with this mass cut,
resulting from varying the fit parameters for the mass spectrum. A pion is
defined as a negative particle with measured mass in the range
 $0. ~-~ 0.4$ $GeV/c^{2}$  
if its measured beta is less than one. Due to our finite time resolution in the
hodoscopes the measured beta of a particle could be greater than one;  any such
particle with negative charge is considered to be a pion.  
When identifying the pion we avoid imposing
strict beta or mass cuts which would imply strict time of flight cuts.
 The reason for this is that both of the decay particles' time of flight measurements have
a common start time. Any fluctuation of the start time will create
correlations in the masses and betas of the $^{3}He$ and $\pi^{-}$ which is magnified by
strict cuts and can create structure in the same event spectrum which is not present
in the mixed event spectrum. 
The mass cut used for the identification of the negative pions is loose and
its efficiency is not significantly different from $100\%$.  
It is also required that with the apparatus divided to two horizontal sections,
the $^{3}He$ be observed in one part of the detector and the $\pi^{-}$
 in the other. The same event invariant mass spectrum (SE) of the $^{3}He$
 nuclei and the negative pions that pass all the cuts
is constructed.  We get the background shape (Bg)
by combining the daughter
particle of one type from one event with the daughter particles of the other type from
the six subsequent events. The linear fit, 
$\alpha\times (Bg)  +  \beta\times (MC)  =  SE$, yields $\beta = 3.37\pm 1.67$.
             
\begin{figure}
\centering\leavevmode\epsfbox{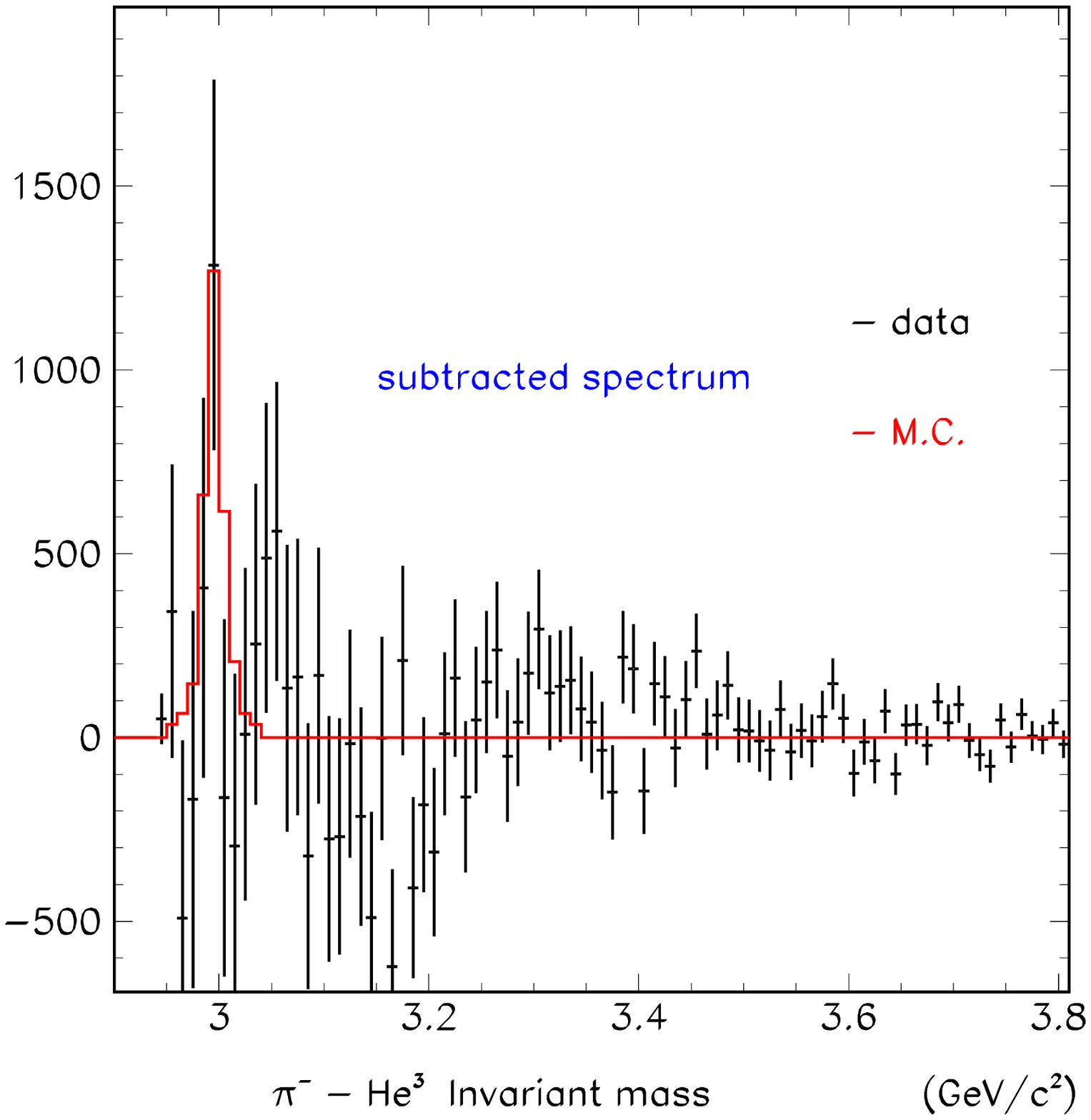}   
\caption{Subtracted invariant mass spectrum for $^{3}He$ - $\pi^{-}$ when
strict hodoscope cuts are applied on the data. The solid histogram overlaid on
the data is
the M.C. $^{3}_{\Lambda}H$ signal normalized so that the peak bin matches the data. }
\label{fig:h3l_0_4}
\end{figure}

The fit parameter $\beta$ suggests a signal
of 2.0 $\sigma$.  Figure~\ref{fig:h3l_0_4} shows
 the subtracted spectrum $SE- \alpha\times Bg$, with
the M.C. shape of the signal overlaid on the data. The $\chi^{2}$ per d.o.f. of the fit
is 1.1 which implies a confidence level of $\sim 32\%$. If we only perform the linear fit
within ten bins of where we expect our signal to be, the signal
to background ratio does not change much and the confidence level of the fit increases
to $\sim 40\%$ which gives us confidence that we actually have a signal.

 Assuming that the peak in the invariant mass spectrum is a signal, the 
invariant yield in a rapidity and $p_{T}$ bin can be calculated. Due to low statistics
the yield has to be calculated in a single rapidity (y: 1.6-2.6) and transverse momentum 
($p_{T}$: 0-1.5GeV/c) bin. The reason that this specific kinematic bin was chosen was that it had reasonable
 acceptance and therefore there is not a large dependence of the yield on the assumed input model, as 
discussed below. Restricting
 the analysis to this kinematic region does however decrease somewhat the significance of the resulting signal.

 The invariant multiplicity of the $^{3}_{\Lambda}H$ is calculated according to:  
\begin{equation}
Y = \frac{1}{2\pi
\bar{p_{T}}\Delta y \Delta p_{T}}\times\frac{N_{count}}{N_{sampled}\times
\epsilon_{total}\times\eta_{eff}}  
\label{eq:yiel}
\end{equation}
where $N_{count}$ is determined by the linear fit parameters
 as $N_{count}=\beta\times (MC)$, 
$N_{sampled}$ is the number of sampled events ($13.5\times 10^{9}$),
 $\Delta y$ is 1, $\Delta p_{T}$ 1.5 GeV/c, $\bar{p_{T}}$ the average transverse momentum
in the $p_{T}$ bin, $\epsilon_{total}$ the geometric acceptance of
$^{3}_{\Lambda}H$ including the LET, ADDMC (reconstruction efficiency), method efficiency and $\eta_{eff}$ the
product of all the other efficiencies. These efficiencies are explained in detail in the following paragraphs.

The geometric acceptance is the
fraction of $^{3}_{\Lambda}H$ nuclei in this kinematic 
bin whose decay products traverse all the 
downstream detectors. The E864
acceptance in rapidity and transverse momentum is determined by
generating a Monte Carlo distribution of  $^{3}_{\Lambda}H$ particles, allowing them to decay
to $^{3}He$ and $\pi^{-}$ and tracking the daughter particles using a full GEANT
simulation of the experiment. The
particle hit information is recorded in each of the detectors and then ``faked'' by
smearing the hits according to the detector resolutions. The ``faked'' data is then 
analysed as the real data but with no cuts besides
the fiducial ones and if both the $^{3}He$ and $\pi^{-}$ traverse all the detectors the
 $^{3}_{\Lambda}H$ rapidity and $p_{T}$ are reconstructed. The geometric acceptance
is the ratio of these accepted $^{3}_{\Lambda}H$ nuclei to the generated ones in the same
kinematic bin. The $^{3}_{\Lambda}H$ nuclei are generated according to a 
production model that is  gaussian in rapidity with   
\begin{equation}
{{dN}\over{dy}} \propto \exp{[-{{(y-y_{cm})^{2}}\over{2\sigma_{y}^{2}}}]}
\label{eq:ninv}
\end{equation}
where $y_{cm}=1.6$ is the center-of-mass rapidity and                          
$\sigma_{y}=1$ which is roughly estimated by folding together the 
rapidity distribution of the ${\Lambda}$  \cite{ahm,barr} and the deuteron 
\cite{arm4}. A Boltzmann distribution is assumed for
the transverse mass with a temperature (inverse slope) of 450 MeV. In order to study the variation of the acceptance with the assumed
production model different widths of the gaussian distribution were used, varying
between 0.7 and 1.1 and slightly different temperatures, including a flat distribution
in $p_{T}$ with $0 \le p_{T} \le 3 $ GeV  giving a
systematic error of $\pm 9\%$.

 The LET efficiency is the fraction
of the particles of a given species that are selected by using a specific LET curve. 
This trigger efficiency can be calculated by applying the LET curve to Monte Carlo
simulated showers that the particles of interest create in the calorimeter. For this
purpose a Monte Carlo $^{3}_{\Lambda}H$ distribution is generated and the $^{3}He$ decay 
daughters that reach the calorimeter are examined. The peak tower energy and time
associated with these $^{3}He$ nuclei are compared with the actual energy-time curve and it is
determined whether the tower fired or not the trigger. The ratio of the number of Monte Carlo 
$^{3}He$
particles that fire the LET to the total number of $^{3}He$ nuclei that reach the calorimeter is the
trigger efficiency. This method has systematic errors that depend on the simulated calorimeter
shower response. Also time shifts in the apparatus will cause a difference between the
real data and the theoretical curve, so a more reliable method of calculating the efficiency,
especially for low efficiencies like these of $^{3}_{\Lambda}H$, is by
using the measured numbers of $^{3}He$ that did and did not fire the trigger. The number
of $^{3}He$ particles firing the trigger is: 

\begin{equation}
N_{LET}=Y\times N_{evt}\times R\times \epsilon_{LET}
\label{eq:let}
\end{equation}
where Y is the production rate of $^{3}He$, $N_{evt}$ the number of LET triggered events, R the
trigger rejection factor, $\epsilon_{LET}$ the trigger efficiency. The number
of $^{3}He$ particles not firing the trigger is: 

\begin{equation}
N_{nonLET}=Y\times N_{evt}\times(1- \epsilon_{LET})
\label{eq:let}
\end{equation}

so

\begin{equation}
\epsilon_{LET}=\frac{N_{LET}}{N_{LET}+R\times N_{nonLET}}
\label{eq:let_cor}
\end{equation}
Since the LET efficiency will be different for runs with different rejection factor R,
it is calculated separately for various runs with different rejection factors 
and the overall efficiency is their weighted average. 

Since it is possible for more than one track to hit the
 same detector element and for these tracks therefore to not be
reconstructed, there is a track reconstruction (or ADDMC) efficiency. The ADDMC efficiency is
determined by using Monte Carlo tracks embedded in real events.

 Finally the method 
efficiency is the efficiency 
of requiring the $^{3}He$ to be in the left part of
the detector and the $\pi^{-}$ in the right.  The method efficiency
 is the ratio of the reconstructed Monte Carlo $^{3}_{\Lambda}H$ in a
kinematic bin after and before the cuts are applied.
 The geometric acceptance of the 
$^{3}_{\Lambda}H$ including the LET, ADDMC and method efficiency 
 ($\epsilon_{total}=\epsilon_{acc}\times\epsilon_{LET}\times\epsilon_{ADDMC}\times
\epsilon_{method}$) 
is listed in Table ~\ref{tab:eff_yi}. Other efficiencies include the detector
efficiency, the charge two cut efficiency, the efficiency of the cuts on the 
$\chi^{2}$ distributions of the various fits applied on the reconstructed tracks, and
the probability that the $^{3}_{\Lambda}H$ will interact with the target.
 All the efficiencies are listed in Table ~\ref{tab:eff_yi}.
Finally only $^{3}_{\Lambda}H$ that are 
reconstructed from $^{3}He$
 that fired the LET can be included in our measured signal. This requirement reduces our counted sample of observed 
$^{3}_{\Lambda}H$ by $84\%$. 
The systematic errors involved in this analysis are the
$9\%$ systematic error in  the calculated acceptance and a $1\%$ systematic error
from the fit to the $^{3}He$ mass peak. 
The $^{3}_{\Lambda}H$ invariant yield, with the statistical
and systematic errors combined, is listed 
in Table ~\ref{tab:eff_yi}.    

As an alternative method of obtaining the $^{3}_{\Lambda}H$ signal, we can 
define the $\pi^{-}$ as a negative
particle with z component of the momentum being less than 3 GeV/c with all other cuts
being kept the same. The effect of the strict
$p_{z}$ cut is estimated by using M.C. data and is very close to$100\%$ efficient. 
We then obtain the fit parameter $\beta = 2.79\pm 1.69$.
 The integrated signal is
 compatible with the results of the first method, as expected by M.C. studies.

Because of the marginal signal, increasing the efficiency 
in detecting the $^{3}_{\Lambda}H$ is important.
One way of achieving this is by requiring that $^{3}He$ and $\pi^{-}$ tracks are separated by a specific distance 
$\left|\Delta x \right|$ in the x direction on each of the detector planes 
instead of requiring that each decay particle is
observed in a different side of the detector. However due to specific considerations that have to
be taken into account to ensure that the same and mixed event spectra are treated
 similarly we
could not achieve a much better signal to background ratio. The results from this
method agree with the results obtained by applying strict hodoscope cuts within the
statistical errors. The invariant yields obtained from the three different
methods are within 25\% of each other.    
The agreement of the results obtained from these various methods make it more
probable that the peak in the invariant mass spectrum is indeed a signal
and not just background fluctuations.

\subsection{$^{4}_{\Lambda}H$ analysis}

\begin{figure}
\centering\leavevmode\epsfbox{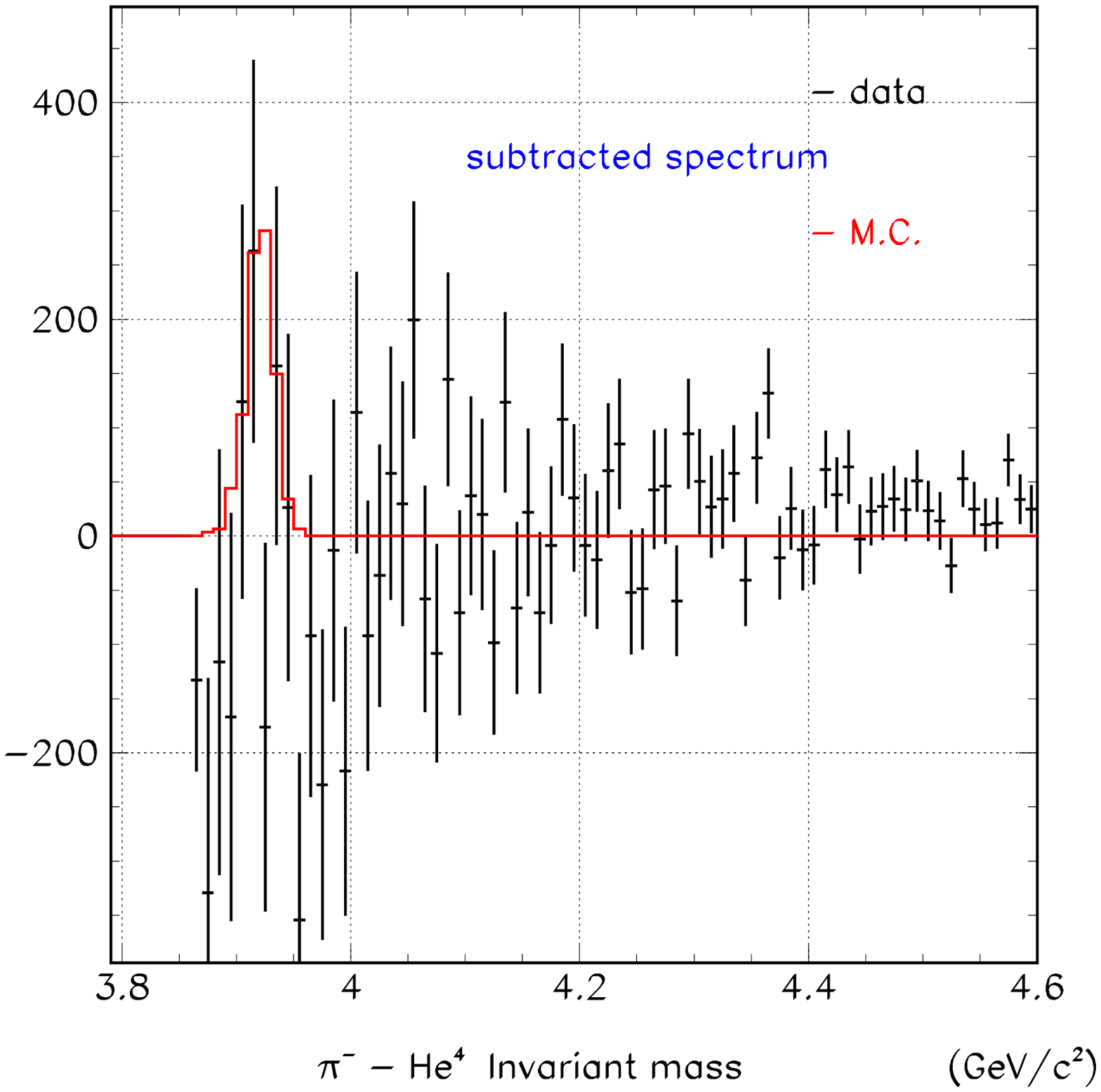}   
\caption{Subtracted invariant mass spectrum for $^{4}He$ - $\pi^{-}$ when
strict hodoscope cuts are applied on the data. The solid histogram overlaid on
the data is
the M.C. $^{4}_{\Lambda}H$ signal }
\label{fig:h4l_0_4}
\end{figure}

The first requirement for the reconstruction of the $^{4}_{\Lambda}H$ signal is to
 identify its
decay products, $^{4}He$ and $\pi^{-}$. A $^{4}He$ is defined as a charge two object
with measured rapidity less than 2.7 and reconstructed mass in the range
 $3.2 ~-~ 6$ $GeV/c^{2}$. The
efficiency of the rapidity cut is incorporated in the geometric acceptance for the 
$^{4}_{\Lambda}H$ and the efficiency of the mass cut is calculated by using real data.
 We
estimate an efficiency of 79\% for the mass cut with a systematic error of 1\%
resulting from varying the parameters of the fit to the mass spectrum.
A pion is
defined as a negative charge object with measured beta  greater than one or a negative
particle with mass in the range $0. ~-~ 0.4$ $GeV/c^{2}$  
if its measured beta is less than one.
Then it is required that the $^{4}He$ is observed in the left part of the detector 
and the $\pi^{-}$ in the right part. The same event invariant mass spectrum (SE) of
 the $^{4}He$ nuclei and the negative
pions  
that pass all the cuts
is constructed  and  we get the background shape (Bg)
by combining the daughter
particle of one type from one event with the daughter particles of the other type from
the six subsequent events. 
The fit parameter $\beta = 0.8\pm 1.4$ suggests that we do not have a statistically 
significant signal and the figure 
~\ref{fig:h4l_0_4} shows the subtracted spectrum $SE - \alpha\times Bg$, 
with the M.C. shape of the signal overlaid on the data.

To estimate a $90\%$ Confidence upper limit of $^{4}_{\Lambda}H$ in the rapidity 
range 1.8 - 2.6 and $p_{T}$ 0 - 1.5
GeV, the fit is performed by using the SE, ME and MC spectrum for pairs with a
reconstructed $^{4}_{\Lambda}H$ rapidity and $p_{T}$ in the above range and it yields
 $\beta = -0.2\pm 1.3$.
The various efficiencies are estimated and listed in Table ~\ref{tab:eff_yi_4}, for
 the rapidity range 1.8 - 2.6 and $p_{T}$ 0 - 1.5 GeV/c.         
The fit parameter $\beta$ is negative which implies the measured signal 
($\beta \times M.C. = -53 \pm 344$) and resultant invariant yield are
unphysical due to random error. The way to deal with that is to calculate
the number Y for the invariant multiplicity according to equation ~\ref{eq:yiel},
 though unphysical, and the corresponding error $\delta Y$ 
by using the fit parameters and from these the gaussian with mean Y and variance
$(\delta Y)^{2}$ is constructed. The physical region of the gaussian is bounded
 from below by 0, so the upper limit $Y_{1}$ is a number such that the integral 
from 0 to $Y_{1}$ is $90\%$ of the integral from 0 to infinity (physical region)
\cite{datb}.
In this way the $90\%$ Confidence Level upper limit is established as $0.42\times 10^{-4}$ $(GeV/c)^{-2}$.

\section{Results and Discussion}
\subsection{ Comparison of Hypernuclei production to non-strange nuclei}

E864 has measured the invariant yields of light stable nuclei with mass number A=1 to
A=7 \cite{arm4}. It is very useful to compare the yields or limits of the light hypernuclei to the
yields of normal nuclei with the same mass number A as it would be very informative
about the coalescence mechanism when strangeness is involved. In particular such a
comparison could indicate an extra penalty factor involved 
in the coalescence of strangeness, if there is any.

In the first place the $^{3}_{\Lambda}H$ is compared to the $^{3}He$ nucleus. Details
of the measurement of $^{3}He$ can be found in \cite{arm4}.
The $^{3}_{\Lambda}H$ yield is approximately a factor of 20 smaller than that of
the $^{3}He$  in the same kinematic region. However in order to make a statement of
whether there is an extra penalty factor when coalescing strangeness the difference in
the production of strange and non-strange baryons should be taken into account.
Therefore the relevant quantity is the ratio 
$\frac{Y_{^{3}_{\Lambda}H}}{Y_{^{3}He}\times\frac{Y_{\Lambda}}{Y_{p}}}$
where $Y_{^{3}_{\Lambda}H}$, $Y_{^{3}He}$, $Y_{\Lambda}$, $Y_{p}$
are the invariant yields of the particles in the rapidity region 1.6 - 2.6 and
$p_{T}$ region  0. - 1.5 GeV/c for $^{3}_{\Lambda}H$ and $^{3}He$  and 0. - 0.5 GeV/c
for the p and $\Lambda$ so as to match velocities. Furthermore the
yields for different species have different kinematic dependences due to collective
motion. Also the efficiency of detecting the $^{3}_{\Lambda}H$, depends strongly on the
rapidity - $p_{T}$ region. The following ratio

\begin{equation}
R \hspace{0.1in} = \frac{Y_{^{3}_{\Lambda}H}}{\sum_{i} 
(Y_{^{3}He}\times\frac{Y_{\Lambda}}{Y_{p}}\times\epsilon)_{i}/
\sum_{i}\epsilon_{i}}
\label{eq:ratio}
\end{equation}
is therefore calculated, where the index i indicates various rapidity and
 $p_{T}$ bins and $\epsilon_{i}$ the
efficiency for detecting the hypernucleus at each bin. Specifically, the momentum space
is divided into bins of 0.5 GeV/c in $p_{T}$ for the $^{3}He$  (0.167 GeV/c for the
protons and $\Lambda$ particles to match velocities) and  0.2 units in rapidity.
                                                       
 The invariant yields for protons
measured by E864 \cite{arm4} are used to calculate the weighted average yields for
 the rapidity, $p_{T}$ bins that are used here. For the 
kinematic regions where there is no measurement for protons,
 we use values obtained by the following parameterization which fits our data:
   
\begin{equation}
Y \propto m_{t}\times exp{[-{{m_{t}-0.938}\over{0.212}}]}
\times[25.8-0.17\times(y-y_{cm})^{2}]
\label{eq:rig_num}
\end{equation}
where $m_{t}$ is the transverse mass in units of $GeV/c^{2}$ . Similarly for the $^{3}He$ the following parameterization

\begin{equation}
Y \propto m_{t}\times exp{[-{{m_{t}-2.809}\over{0.405}}]}
\times[8+17\times(y-y_{cm})^{2}]
\label{eq:rig_num}
\end{equation}
is used for the bins where there was no measurement.
 Finally for the $\Lambda$ the results
of experiment E891 and E877 were used and the production model assumed \cite{ahm} was

\begin{equation}
Y \propto exp{[-m_{t}\times(4.3 + 6.5\times cosh(y-y_{cm}) 
- 4.2\times(y-y_{cm})^{2})]}.
\label{eq:rig_num}
\end{equation}

The value obtained for the ratio of Equation~\ref{eq:ratio} is 
\begin{equation}
R \hspace{0.1in} = \frac{Y_{^{3}_{\Lambda}H}}{\sum_{i} 
(Y_{^{3}He}\times\frac{Y_{\Lambda}}{Y_{p}}\times\epsilon)_{i}/
\sum_{i} \epsilon_{i}}=0.36\pm0.26
\label{eq:ratio_re}
\end{equation}
which indicates that there is an extra suppression in the coalescence 
production of 
$^{3}_{\Lambda}H$ of about a factor of three compared to that of $^{3}He$, after accounting for the different abundances of the
coalescence ingredients.
   
In the case of $^{4}_{\Lambda}H$ the $90\%$ upper limit can be compared to the yield of
$^{4}He$. In E864 the $^{4}He$ production has been measured \cite{arm4}
 and the following parameterization is used for the regions where
there is no measurement
\begin{equation}
Y \propto m_{t}\times exp{[-{{m_{t}-3.727}\over{0.435}}]}
\times[1+3.17\times(y-y_{cm})^{2}]
\label{eq:rig_num}
\end{equation}
                                                       
The relevant ratio for the $^{4}_{\Lambda}H$ is
\begin{equation}
R \hspace{0.1in} = \frac{Y_{^{4}_{\Lambda}H}}{\sum_{i} 
(Y_{^{4}He}\times\frac{Y_{\Lambda}}{Y_{p}}\times\epsilon)_{i}/
\sum_{i} \epsilon_{i}}
\label{eq:ratio_h4l}
\end{equation}
where $\epsilon_{i}$ is the efficiency for detecting the $^{4}_{\Lambda}H$ and 
$Y_{^{4}_{\Lambda}H}$ the (unphysical) invariant yield for the $^{4}_{\Lambda}H$.
The ratio R and the relevant error of the ratio $\delta R$ are calculated and the 
gaussian with mean R and variance
$(\delta R)^{2}$ is constructed. The physical region of the gaussian is bounded
from below by 0, so the $90\%$ confidence upper limit $R_{1}$ is a number such that the integral 
from 0 to $R_{1}$ is $90\%$ of the integral from 0 to infinity (physical region).
 Additionally the fact that 
$^{4}_{\Lambda}H$ has a ground state with spin J=0 and an excited state with spin J=1, 
which decays to the ground state, 
whereas the $^{4}He$ has only a ground state of J=0 has to be taken into account since
the invariant yields of different species are proportional to their spin degeneracy
factors (2J+1) \cite{arm4}. So the spin-corrected $90\%$ Confidence Level upper limit of the
ratio is established as
  
\begin{equation}
R_{2} \hspace{0.1in} =\frac{1}{4}\times R_{1}\leq 0.225
\label{eq:ratio_h4l2}
\end{equation}
which indicates an extra suppression in the production of 
$^{4}_{\Lambda}H$ of at least a factor of four as compared to that of $^{4}He$.

\subsection{ Implications of $^{3}_{\Lambda}H$, $^{4}_{\Lambda}H$  results}

There is an apparent suppression in the
production of the light hypernuclei as compared with the yields of non - strange
nuclei. However before making any conclusions about an extra penalty factor involved
when coalescing strangeness, a number of other possible reasons for this suppression
should be examined. In particular both hypernuclei are very weakly bound and therefore
could be easily destroyed in final state soft interactions, or their relatively large
size could be a factor in their production.

\subsubsection{ Finite Size Effects}

The effect of the finite size of nuclear clusters in their production has been
 studied by Scheibl and Heinz \cite{hei}.
Their coalescence model includes the dynamical
expansion of the collision zone which results in correlations
 between the momenta and
positions of the particles at freeze-out. The invariant spectrum of the formed clusters
with mass number A and transverse mass $M_{t}$  
is proportional to some effective volume $V_{eff}(A, M_{t})$. In the saddle - point 
approximation this effective volume is
proportional to the ``homogeneity volume'' $V_{hom}(m_{t})$ of the constituent nucleons 
with transverse mass $m_{t}$ and transverse momentum $p_{T}$, where
$p_{T}=p_{T}/A$ with $p_{T}$ being the transverse momentum of the nucleus. The 
$V_{hom}(m_{t})$ is accessible through HBT measurements.  

Nuclear clusters cannot be treated as pointlike particles since
their rms radii are not much smaller than the homogeneity radii.
 A quantum mechanical
density matrix approach allows the inclusion of finite size effects and the
 internal
cluster wave function \cite{hei}. The number of created clusters at a specific momentum is
given by projecting the cluster's density matrix onto the constituent nucleons density
matrices in the fireball at freeze-out. In the case of the deuteron one of the internal wave
functions considered is the spherical harmonic oscillator with size parameter d=3.2 fm.
Under various assumptions it is shown that the yield of the deuterons is identical
with the classic thermal spectrum with only an extra quantum mechanical correction
factor $C_{d}(R_{d},P_{d})$, where $R_{d}$ and $P_{d}$ are the deuteron's
 momentum and center of mass 
space-time coordinates in the fireball rest frame. $C_{d}$ provides a measure for the
homogeneity of the nucleon phase-space around the deuteron center-of-mass coordinates.
The measured deuteron momentum spectra do not contain information on the point of
formation, so the average correction factor over the freeze-out hypersurface is the
relevant quantity, which has a simple approximate form:
\begin{equation}
<C_{d}>\hspace{0.1in}\approx \frac{1}{(1+(\frac{d}{2R_{T}(m)})^{2})\times \sqrt{1+
(\frac{d}{2R_{||}(m)})^{2}}}
\label{eq:cor_fa}
\end{equation}
where $R_{||}(m_{t}),R_{T}(m_{t})$
are the longitudinal and transverse lengths of homogeneity for the constituent
nucleons.                                                    
This means that the approximate average correction factor only depends on the ratio of
the size parameter of the deuteron's wave function d to the radii of homogeneity of the
constituent nucleons with zero transverse momentum.

The $^{3}_{\Lambda}H$ has an rms radius of $\approx 5 fm$ \cite{nem} which is much bigger than the
rms radius of $^{3}He$ (1.74 fm). It is therefore important to investigate the effect
of these different sizes to the relevant yields of the nuclei. In order to make a rough
approximation, it is assumed that the $^{3}_{\Lambda}H$ is a system similar to the
deuteron that consists of a $\Lambda$ and a $^{2}H$ and the $^{3}He$ of a proton and a
$^{2}H$. If a harmonic oscillator
internal wave function is assumed for the systems with a size parameter for 
$^{3}_{\Lambda}H$ equal to the
mean distance of the $\Lambda$ to the c.m. of the $^{2}H$,
 $d_{^{3}_{\Lambda}H}=\sqrt{<r^{2}>_{\Lambda d}}\approx 9.8 fm$, 
and for the $^{3}He$ equal to the mean distance of a proton to the c.m. of the 
$^{2}H$, $d_{^{3}He}=\sqrt{<r^{2}>_{p d}}\approx 2.6 fm$,
then the relevant correction factors can be calculated by using equation ~\ref{eq:cor_fa}.
Various AGS experiments have estimated the radii of homogeneity \cite{barr1,barr2,lis,mis,fil} varying from 
3 to 10 fm. By substituting in equation ~\ref{eq:cor_fa} we obtain 
a rough approximation of 
the two correction factors.
\begin{equation}
\frac{C_{^{3}_{\Lambda}H}}{C_{^{3}He}} \approx 0.41\pm 0.1
\label{eq:v}
\end{equation}
which implies that the yield of $^{3}_{\Lambda}H$ as compared to that of $^{3}He$ could
be a factor between two and three smaller just due to size effects.

In the case of the $^{4}_{\Lambda}H$, its rms radius of 2 fm \cite{nem} is not much
 bigger than
the $^{4}He$ radius (1.41 fm). Following a similar treatment as before, where the 
$^{4}_{\Lambda}H$
is considered to be a $\Lambda$ bound to a triton and $^{4}He$ a proton bound to a
triton, with size parameters, 
$d_{^{4}_{\Lambda}H}=\sqrt{<r^{2}>_{\Lambda t}}\approx 3.9 fm$ and 
$d_{^{4}He}=\sqrt{<r^{2}>_{p t}}\approx 1.9 fm$
we have
\begin{equation}
\frac{C_{^{4}_{\Lambda}H}}{C_{^{4}He}} \approx \frac{1}{1.15}.
\label{eq:v}
\end{equation}
The size effect is not so important in the case of the $^{4}_{\Lambda}H$. 

\subsubsection{ Small Binding Energy}

We have previously reported that the yields of light nuclei near midrapidity and at low 
$p_{T}$ are well described by the formula $(\frac{1}{48})^{A} \times exp[-B/T_{s}]$
 where B is the
binding energy per nucleon and inverse slope $T_{s}=5.9\pm 1.1 MeV$ \cite{arm5}.
 This dependence cannot be explained by the coalescence or thermal models that assume a
simple exponential dependence on the total binding energy B of the form  exp[-B/T] 
where the temperature T
of the collisions at freeze-out are in the order of 100-140 MeV. 

The binding energy per nucleon in the case of the $^{3}_{\Lambda}H$ is 0.8 MeV whereas
for the $^{3}He$ it is  2.7 MeV. Therefore just by taking into account the exponential 
dependence described above the $^{3}_{\Lambda}H$ yield should be only 70\% of that of
$^{3}He$.

In the case of the $^{4}_{\Lambda}H$ which has a binding energy per nucleon of 2.5 MeV
and the $^{4}He$ with 7 MeV/nucleon the effect is even bigger. The $^{4}_{\Lambda}H$
yield is suppressed by an extra factor of 2.2 relative to the $^{4}He$ yield. 

When examining the effect of the binding energy  in the production of
hypernuclei, it would be informative to compare the $^{4}_{\Lambda}H$ yield to 
that of the particle unstable nucleus 
$^{4}H$. In E864 the yields of particle-unstable light nuclei have been measured 
\cite{arm6}.
The invariant yields of $^{4}H$ are within 50\% of the $^{4}He$ yields,
 even though the $^{4}He$ has only a
ground state of spin J=0 whereas the spin of $^{4}H$ is two (J=2) which
means that the $^{4}H$ yields should be a factor of five greater than those of
$^{4}He$. If we assume feeddown from the excited states of $^{4}H$ (J=1,0,1)
then the yields should be about an order of magnitude higher. Again it is not
 possible to explain this apparent suppression in the realm of the thermal and
coalescence models since the mass difference
between the stable and unstable nuclei is small compared to the collision temperatures
$(\approx 100 MeV)$. The spin-corrected ratio for the $^{4}_{\Lambda}H$ to $^{4}H$ is

\begin{equation}
R \hspace{0.1in} =\frac{10}{4}\times \frac{Y_{^{4}_{\Lambda}H}}{\sum_{i} 
(Y_{^{4}H}\times\frac{Y_{\Lambda}}{Y_{n}}\times\epsilon)_{i}/
\sum_{i} \epsilon_{i}}\leq 1.4
\label{eq:ratio_fin}
\end{equation}
which does not indicate any extra suppression in the production of 
$^{4}_{\Lambda}H$ as compared to that of $^{4}H$. Of course this cannot be conclusive for the existence or not of any extra
suppression as we just have an upper limit for the $^{4}_{\Lambda}H$.

\section{Summary}

There is an apparent suppression in the production levels of  $^{3}_{\Lambda}H$ and 
 $^{4}_{\Lambda}H$  as compared with the stable light nuclei of the same mass number.
However the size of  $^{3}_{\Lambda}H$ and the small binding Energy of both hypernuclei
seems to be an important factor in determining their yields. Therefore
there is no clear indication of an extra penalty factor involved in the coalescence of
strangeness and future studies are important to clarify these coalescence factors.

\section{Acknowledgements}

We gratefully acknowledge the efforts of the AGS staff in providing the beam.  
This work was supported in part by grants from the Department of Energy 
(DOE) High Energy Physics Division, the DOE Nuclear Physics Division, and 
the National Science Foundation.


\newpage

%
%

%
%
\newpage

\begin{table}
\begin{tabular}{|c|c|}
rapidity &  1.6 - 2.6          \\
 $p_{T}$  (GeV/c)   &    0 - 1.5           \\ \tableline \tableline
$N_{event}$           & $13.5\times 10^{9}$        \\ \tableline 
$N_{count}$           & $1220\pm854$        \\ \tableline 
$\epsilon_{total}=\epsilon_{acc}\times\epsilon_{LET}\times\epsilon_{ADDMC}\times
\epsilon_{method}$           &$1.96\times 10^{-4}$       \\ \tableline 
detector efficiency: $\eta_{det}$           & $0.82^{2}$       \\ \tableline 
charge cut efficiency: $\eta_{q}$           & 0.84       \\ \tableline 
target absorption probability: $\eta_{targ}$           & 0.78       \\ \tableline 
$\chi^{2}$ cut efficiency: $\eta_{\chi^{2}}$           & 0.90       \\ \tableline  
invariant yield $(GeV/c)^{-2}$   & $(5.27\pm 4.04)\times10^{-4}$ \\ \tableline   
\end{tabular}
\caption{Efficiencies and invariant yields for $^{3}_{\Lambda}H$ in $10\%$ most central
Au-Pt collisions in units of $ c^{2}/GeV^{2}$ for y:1.6-2.6 and $p_{T}$: 0-1.5 GeV/c  }
\label{tab:eff_yi}
\end{table}  

\begin{table}
\centering
\begin{tabular}{|c|c|}
\hline
rapidity            & 1.8 - 2.6          \\
 $p_{T}$  (GeV/c)            & 0 - 1.5           \\ \tableline \tableline
$N_{event}$           & $13.5\times 10^{9}$        \\ \tableline 
$\epsilon_{total}=\epsilon_{acc}\times\epsilon_{LET}\times\epsilon_{ADDMC}\times
\epsilon_{method}$           &$9.3\times 10^{-4}   $       \\ \tableline 
detector efficiency: $\eta_{det}$           & $0.82^{2}$       \\ \tableline  
charge cut efficiency: $\eta_{q}$           & 0.84       \\ \tableline  
target absorption probability: $\eta_{targ}$           & 0.89       \\ \tableline  
$\chi^{2}$ cut efficiency: $\eta_{\chi^{2}}$           & 0.90       \\  
\end{tabular}
\caption{Efficiencies for the $^{4}_{\Lambda}H$ in $10\%$ most central
Au-Pt collisions in units of $ c^{2}/GeV^{2}$ for y:1.8-2.6 and $p_{T}$: 0-1.5 GeV/c  }
\label{tab:eff_yi_4}
\end{table}

\end{document}